\documentclass{icrc29_astro}
\usepackage{graphicx,amssymb,amsmath,times}
\newcommand{\rsub}{r_\mathrm{sub}}
\newcommand{\xx}[1]{\times 10^{#1}}
\newcommand{\TDS}{T_\mathrm{DS}}
\newcommand{\pth}{p_\mathrm{th}}
\newcommand{\pinj}{p_\mathrm{inj}}
\newcommand{\pmax}{p_\mathrm{max}}

\newcommand{\alf}{Alfv\'en}

\newcommand{\vthres}{v^{\mathrm{MC}}_\mathrm{thres}}

\newcommand{\injSA}{\xi_\mathrm{SA}}
\newcommand{\etaSA}{\eta_\mathrm{SA}}
\newcommand{\etaMC}{\eta_\mathrm{MC}}
\newcommand{\SA}{semi-analytic}

\newcommand{\MB}{Maxwell-Boltzmann}
\newcommand{\MC}{Monte Carlo}
\newcommand{\mc}{Monte Carlo}

\newcommand{\fofp}{f(p)}

\newcommand{\NL}{nonlinear}

\newcommand{\rel}{relativistic}

\newcommand{\ultrarel}{ul\-tra-rel\-a\-tiv\-is\-tic}
\newcommand{\TP}{test-particle}

\newcount\listno
\def\List{\global\advance \listno by 1 {(\the\listno)}}

\newcount\listcno
\def\Listc{\global\advance \listcno by 1 
	{({\expandafter{\romannumeral\listcno})\,}}}
	\def\newlistc{\listcno=0}

\begin{document}
\title[Injection in
Nonlinear Diffusive Shock Acceleration]{Thermal Particle Injection in
Nonlinear Diffusive Shock Acceleration}
\author[Ellison, Blasi \& Gabici] {Donald C. Ellison$^a$, Pasquale Blasi$^b$, 
 Stefano Gabici$^c$ \\
        (a) Physics Dept., North Carolina State Univ., Raleigh, NC 27695, U.S.A. \\ 
        (b) High Energy Astrophysics Group, INAF,
        Osservatorio Astrofisico di Arcetri, Largo E. Fermi 5, I-50125, Firenze,
        Italy \\
        (c) Humboldt Fellow, Max-Planck-Institut fuer Kernphysik, Saupfercheckweg 1,
69117 Heidelberg, Germany }
\presenter{Presenter: Don Ellison (don$\_$ellison@ncsu.edu), \  
usa-ellison-D-abs3-og14-oral}

\maketitle

\begin{abstract}
Particle acceleration in collisionless astrophysical shocks, i.e., diffusive shock
acceleration (DSA), is the most likely mechanism for producing cosmic rays, at least
below $10^{15}$ eV. Despite the success of this theory, several key elements,
including the injection of thermal particles, remains poorly understood. We
investigate injection in strongly nonlinear shocks by comparing a semi-analytic model
of DSA with a Monte Carlo model. These two models treat injection quite differently
and we show, for a particular set of parameters, how these differences influence the
overall acceleration efficiency and the shape of the broad-band distribution
function.
%
%
\end{abstract} \vskip-6pt

\vskip-24pt\hbox{}
\section{Introduction} \vskip-6pt
We compare a recent semi-analytic (SA) model of non-linear diffusive shock
acceleration (DSA) \cite{Blasi2002}\cite{Blasi2004}\cite{BGV2005} with a
well-established Monte Carlo (MC) model (e.g.,
\cite{EMP90}\cite{JE91}\cite{EJB99}). Both include a thermal leakage model for
injection and the non-linear (NL) backreaction of accelerated particles on the shock
structure, but they do these in very different ways. Also different is the way in
which the particle diffusion in the background magnetic turbulence is modeled.
%
%
Our limited comparison shows that the important NL effects of compression ratios $\gg
4$ and concave spectra do not depend strongly on the injection model as long as
injection is efficient. A fuller understanding of the complex plasma processes
involved, particularly if injection is weak, will require particle-in-cell (PIC)
simulations (e.g., \cite{GE2000}), but these simulations, which must be done fully in
three-dimensions\cite{JJB98}, cannot yet be run long enough, in large enough
simulation spaces, to accelerate particles from thermal to \rel\ energies in order to
show strong NL effects. For now, approximate methods must be used.

The \MC\ model is more general than the semi-analytic model,\footnote{For
instance, the MC model can treat a specific momentum dependence for the scattering
mean free path, particle
acceleration in \rel\ shocks\cite{ED02}\cite{ED04}, and NL effects in oblique
shocks\cite{EJB99}.} but it is considerably slower computationally.
Since it is important in many applications, such as hydro models of supernova
remnants, to include the dynamic effects of \NL\ DSA in simulations which perform the
calculation many times (e.g., \cite{EDB2004}), a rapid, approximate calculation, such
as the SA one discussed here, is useful.

\vskip-24pt\hbox{}
\section{Models} \vskip-6pt
%
The main features of the two models that are relevant for our comparison here are:

{\bf Injection:}\ In the \SA\ model, a free injection parameter, $\injSA$, determines
the fraction of total particles injected into the acceleration mechanism and the
injection momentum, $\pinj$. Specifically, $\pinj = \injSA\, \pth$, where $\pth =
\sqrt{2 m_p k \TDS}$ ($\TDS$ is the downstream temperature). The fraction, $\etaSA$,
of unshocked particles crossing the shock which become superthermal in the SA model
is
$
\etaSA = (4 / 3 \pi^{1/2}) (\rsub -1) \injSA^3 e^{-\injSA^2}
$,
%
%
%
%
where $\rsub$ is the subshock compression ratio.
The fraction $\etaSA$, which is approximately the number of particles in the
Maxwellian defined by $\TDS$ with momentum $p> \pinj$, is determined by requiring the
continuity, at $\pinj$, of the Maxwellian and the superthermal distribution (see
\cite{BGV2005}).  Since $\pth$ depends on the injected fraction, the solution must be
obtained by iteration.

In the \mc\ model, the injection depends on the scattering assumptions. 
We assume that particles pitch-angle scatter elastically and isotopically in the
local plasma frame and that the mean free path (mfp) is proportional to the
gyroradius, i.e., $\lambda \propto r_g$, where $r_g = pc/(qB)$.
%
%
With these assumptions, the injection is purely statistical with those ``thermal''
particles which manage to diffuse back upstream gaining additional energy and
becoming superthermal. Note that in this scheme the viscous subshock is assumed to be
transparent to all particles, even thermal ones, and that any downstream particle
with $v \ge u_2$ has a chance to be injected. Here, $u_2$ is the downstream flow
speed.

For comparison with the SA model, we have included an additional parameter,
$\vthres$, to limit injection in the MC simulation. Only downstream particles with
$v\ge \vthres$ are injected, i.e., allowed to re-cross the shock into the upstream
region and become superthermal. In our previous \mc\ results, with the sole exception
of \cite{Ellison85}, we have taken $\vthres=0$.

{\bf Momentum dependent diffusion:}\ Diffusion is treated very differently in the two
models. As just mentioned, the MC simulation models pitch-angle diffusion by assuming
a $\lambda(p)$ 
(see \cite{ED04}). 
The \SA\ model does not explicitly describe diffusion but assumes only that the
diffusion is a strongly increasing function of particle momentum $p$ so that
particles of different $p$ interact with different spatial regions of the upstream
precursor. Eichler \cite{Eichler84} used a similar procedure. With this assumption,
particles of momentum $p$ can be assumed to feel some average precursor fluid speed
$u_p$ and an average compression ratio $r_p \sim u_p/u_2$ (see \cite{BGV2005}).

\begin{figure}[ht]
\begin{center}
\includegraphics*[width=0.6\textwidth,angle=0,clip]{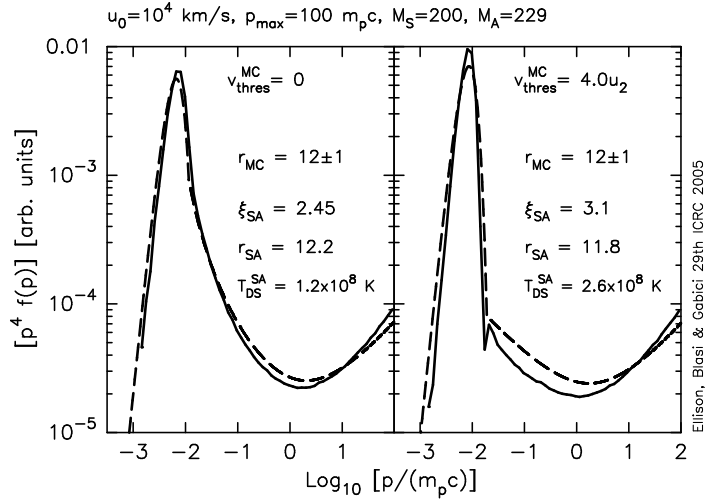}
\caption{\label {fig1} Phase space distributions $\fofp$ times $p^4$ for a MC model
  (solid curves) and a SA model (dashed curves). In the left panel $\vthres=0$, and
  in the right panel $\vthres=4u_2$ in the MC model. Here, $u_2 = u_0/r$ is the
  downstream plasma speed in the shock frame, $u_0$ is the shock speed, and the Mach
  numbers, compression ratios, and shocked temperatures are indicated. In all cases,
  the shock is parallel and \alf\ heating
  is assumed in the shock precursor \cite{EBB2000}. }
\end{center}
\end{figure} \vskip-12pt

The different way diffusion is treated influences not only injection, but also the
shape of the distribution function $\fofp$ [$\fofp$ is the momentum phase space
density, i.e., particles/(cm$^3$-$d^3p$)].
Both models give the characteristic concave $\fofp$ which hardens with increasing $p$
and, since the overall shock compression ratio $r$ can be greater than 4, this
spectrum will be harder than $p^{-4}$ at \ultrarel\ energies. In the results we show
here, the acceleration is limited with a cutoff momentum so $\fofp$ cuts off abruptly
at $\pmax$.  More realistic models will show the effects of escape from some spatial
boundary (e.g., finite shock size) or from a finite acceleration time. In either
case, the spectrum will show a quasi-exponential turnover, e.g., $\fofp \propto
p^{-\sigma} \exp{[-\alpha^{-1} (p/\pmax)^{\alpha} ]}$ \cite{EBB2000}, where $\alpha$
is included to emphasize that the detailed shape of the turnover depends on the
momentum dependence of the diffusion coefficient near $\pmax$.

{\bf Thermalization:}\ There is no thermalization process in the MC simulation in the
sense of particles exchanging energy between one another because particles scatter
elastically in the local frame. However, a quasi-thermal low energy distribution is
created as unshocked particles cross the shock from upstream to downstream at
different angles and receive different fractions of the speed difference, $u_0 -
u_2$.  For the parameters used here, the low-energy peak is essentially a Maxwellian
when $\vthres=0$.

In the \SA\ model, the shocked thermal pressure and density are determined from the
conservation relations and these are translated to a \MB\ distribution.

\begin{figure}[ht]
\begin{center}
\includegraphics*[width=0.6\textwidth,angle=0,clip]{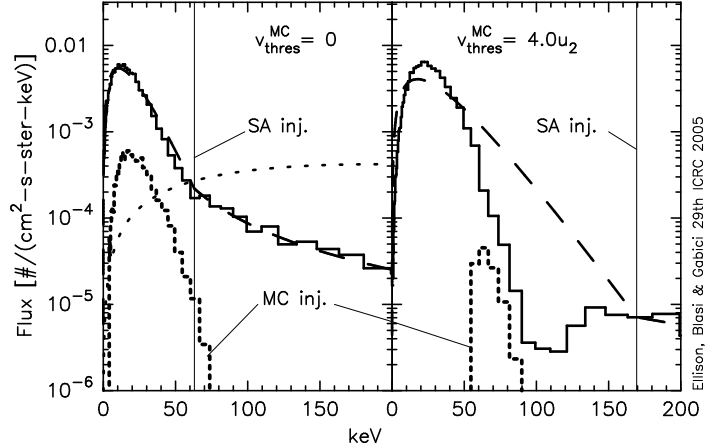}
\caption{\label {fig2} Low energy portions of the spectra shown in
  Fig.~\ref{fig1} where the solid histograms are the MC results and the dashed curves
  are the SA results. The heavy dotted histograms show the distribution of particles
  that were injected in the MC model. The solid vertical lines are drawn at $\pinj$,
  the momentum at which particles are injected in the SA model. For $\vthres=0$,
  $\etaMC=5.7\xx{-2}$ and $\etaSA=4.6\xx{-2}$, while for $\vthres=4u_2$,
  $\etaMC=4.1\xx{-3}$ and $\etaSA=2.5\xx{-3}$. The light-weight dotted curve shows
  the Maxwellian ($\TDS=2.2\xx{9}$~K) that would have resulted if no DSA occurred.}
\end{center}
\end{figure} \vskip-12pt

\vskip-24pt\hbox{}
\section{Results} \vskip-6pt
In Fig.~\ref{fig1} we show distribution functions (times $p^4$) for a set of
parameters and with $\vthres=0$ (left panel) and $\vthres=4 u_2$ (right panel). In
both panels, the solid curves are the MC results and the dashed curves are the SA
results where $\injSA$ has been chosen to provide the best match to the MC
spectra.\footnote{Note that once the input parameters (i.e., $u_0$, $\pmax$,
$\vthres$, and the sonic and \alf\ Mach numbers: $M_S$ and $M_A$) are fixed, the only
parameter varied to match the spectra shown in Figs.~\ref{fig1} and \ref{fig2} is
$\injSA$. The normalization is not adjusted.}
In Fig.~\ref{fig2} we show the low energy portions of the spectra along with the
distributions of injected particles in the MC model (dotted histograms). The vertical
solid lines are at $\pinj$ and they show the transition between the Maxwellian and
the superthermal population in the SA results.

\newlistc

The important aspects of the plots are:

\vskip-3pt
{\Listc}\ The broad-band match between the two very different calculations is quite
good, particularly for $\vthres=0$. Both models show the important characteristics
of \NL\ DSA, i.e., $r \gg 4$, concave spectra, and a sharp reduction in the
shocked temperature from \TP\ values.

\vskip-3pt
\ {\Listc}\ The shocked temperature depends on $\vthres$ with weaker injection
(i.e., larger $\vthres$) giving a larger $\TDS$.\footnote{The light-weight dotted
curve in the left panel of Fig.~\ref{fig2} is a Maxwellian at $\TDS=2.2\xx{9}$~K, the
temperature of the shocked gas without DSA.}

\vskip-3pt
\ {\Listc}\ The distribution $\fofp$ is harder near $\pmax$ in the MC
results than with the SA calculations.  
A common prediction of SA models is that the shape of the particle spectra at $\pmax$
has the form $p^{-3.5}$ if the shock is strongly modified and the diffusion
coefficient grows fast enough in momentum. This can be demonstrated by solving the
equations in the extreme case of a maximally modified shock and approximating the
spectrum with a power law at high momenta. The MC model makes no such power-law
assumption.

\vskip-3pt
\ {\Listc}\ The minimum in the $p^4 \fofp$ plot occurs at $p>m_p c$ in both
models. The transition between $\fofp$ softer than $p^{-4}$ and $\fofp$ harder than
$p^{-4}$ varies with shock parameters and increases as $\pmax$ increases. This is an
important difference from the algebraic model of Berezhko \& Ellison\cite{BE99} where
the minimum is fixed at $m_p c$.

\vskip-3pt
\ {\Listc}\ For the particular parameters used in these examples, the overall
compression ratio is relatively insensitive to $\vthres$, but shocks having other
parameters may show a greater sensitivity. Note that \alf\ wave heating is
assumed in all of the results presented here. If only adiabatic heating was assumed,
the compression ratios would be much higher (see \cite{EBB2000}).

\vskip-3pt
\ {\Listc}\ In all cases, the MC model injects more particles than the SA model
but the average energy of the injected
particles is less, as indicated by the peak of the curve labelled `MC inj' vs. the
`SA inj' energy in Fig.~\ref{fig2}.

\vskip-3pt
\ {\Listc}\ In contrast to $\vthres=0$, the `thermal' part of $\fofp$ with
$\vthres=4u_2$ (Fig.~\ref{fig2}) shows large differences in the two models. While
both conserve particle, momentum, and energy fluxes so that the broad-band $\fofp$
matches well for a wide range of parameters, the different treatments of the subshock
lead to large differences in the critical energy range $2 \lesssim [E/(k\TDS)]
\lesssim 5$. This offers a way to distinguish these models observationally.

\vskip-24pt\hbox{}
\section{Discussion and Conclusions} \vskip-6pt
Using two approximate acceleration models, we have shown that the most important
features of NL DSA, i.e., $r\gg 4$ and concave spectra, are robust and do not
strongly depend on the injection model as long as injection is efficient.  If
injection is weak, as might be the case in highly oblique shocks, accelerated spectra
will depend more on the details of injection, at least in the transition range between
thermal and superthermal energies.
Also, the relative efficiencies for injecting and accelerating electrons vs. protons
or protons vs. heavier ions may require a more detailed description of injection, as
may be provided by future PIC simulations.


{\bf Acknowledgements:} D.C.E. wishes to acknowledge support from a NASA grant
(ATP02-0042-0006) and S.G. acknowledges support from the Humboldt foundation.

\newcommand\itt{ }
\newcommand\bff{ }
\newcommand{\aaDE}[3]{ 19#1, A\&A, #2, #3}
\newcommand{\aatwoDE}[3]{ 20#1, A\&A, #2, #3}
\newcommand{\aatwopress}[1]{ 20#1, A\&A, in press}
\newcommand{\aasupDE}[3]{ 19#1, {\itt A\&AS,} {\bff #2}, #3}
\newcommand{\ajDE}[3]{ 19#1, {\itt AJ,} {\bff #2}, #3}
\newcommand{\anngeophysDE}[3]{ 19#1, {\itt Ann. Geophys.,} {\bff #2}, #3}
\newcommand{\anngeophysicDE}[3]{ 19#1, {\itt Ann. Geophysicae,} {\bff #2}, #3}
\newcommand{\annrevDE}[3]{ 19#1, {\itt Ann. Rev. Astr. Ap.,} {\bff #2}, #3}
\newcommand{\apjDE}[3]{ 19#1, {\itt ApJ,} {\bff #2}, #3}
\newcommand{\apjtwoDE}[3]{ 20#1, {\itt ApJ,} {\bff #2}, #3}
\newcommand{\apjletDE}[3]{ 19#1, {\itt ApJ,} {\bff  #2}, #3}
\newcommand{\apjlettwoDE}[3]{ 20#1, {\itt ApJ,} {\bff  #2}, #3}
\newcommand{\apjpress}{{\itt ApJ,} in press}
\newcommand{\apjletpress}{{\itt ApJ(Letts),} in press}
\newcommand{\apjsDE}[3]{ 19#1, {\itt ApJS,} {\bff #2}, #3}
\newcommand{\apjstwoDE}[3]{ 19#1, {\itt ApJS,} {\bff #2}, #3}
\newcommand{\apjsubDE}[1]{ 19#1, {\itt ApJ}, submitted.}
\newcommand{\apjsubtwoDE}[1]{ 20#1, {\itt ApJ}, submitted.}
\newcommand{\appDE}[3]{ 19#1, {\itt Astropart. Phys.,} {\bff #2}, #3}
\newcommand{\apptwoDE}[3]{ 20#1, {\itt Astropart. Phys.,} {\bff #2}, #3}
\newcommand{\araaDE}[3]{ 19#1, {\itt ARA\&A,} {\bff #2},
   #3}
\newcommand{\assDE}[3]{ 19#1, {\itt Astr. Sp. Sci.,} {\bff #2}, #3}
\newcommand{\grlDE}[3]{ 19#1, {\itt G.R.L., } {\bff #2}, #3} 
\newcommand{\icrcplovdiv}[2]{ 1977, in {\itt Proc. 15th ICRC (Plovdiv)},
   {\bff #1}, #2}
\newcommand{\icrcsaltlake}[2]{ 1999, {\itt Proc. 26th Int. Cosmic Ray Conf.
    (Salt Lake City),} {\bff #1}, #2}
\newcommand{\icrcsaltlakepress}[2]{ 19#1, {\itt Proc. 26th Int. Cosmic Ray Conf.
    (Salt Lake City),} paper #2}
\newcommand{\icrchamburg}[2]{ 2001, {\itt Proc. 27th Int. Cosmic Ray Conf.
    (Hamburg),} {\bff #1}, #2}
\newcommand{\JETPDE}[3]{ 19#1, {\itt JETP, } {\bff #2}, #3}
\newcommand{\jgrDE}[3]{ 19#1, {\itt J.G.R., } {\bff #2}, #3}
\newcommand{\jgrtwoDE}[3]{ 20#1, {\itt J.G.R., } {\bff #2}, #3}
\newcommand{\mnrasDE}[3]{ 19#1, {\itt MNRAS,} {\bff #2}, #3}
\newcommand{\mnrastwoDE}[3]{ 20#1, {\itt MNRAS,} {\bff #2}, #3}
\newcommand{\mnraspress}[1]{ 20#1, {\itt MNRAS,} in press}
\newcommand{\natureDE}[3]{ 19#1, {\itt Nature,} {\bff #2}, #3}
\newcommand{\naturetwoDE}[3]{ 20#1, {\itt Nature,} {\bff #2}, #3}
\newcommand{\nucphysA}[3]{#1, {\itt Nuclear Phys. A,} {\bff #2}, #3}
\newcommand{\pfDE}[3]{ 19#1, {\itt Phys. Fluids,} {\bff #2}, #3}
\newcommand{\phyreptsDE}[3]{ 19#1, {\itt Phys. Repts.,} {\bff #2}, #3}
\newcommand{\physrevEDE}[3]{ 19#1, {\itt Phys. Rev. E,} {\bff #2}, #3}
\newcommand{\prlDE}[3]{ 19#1, {\itt Phys. Rev. Letts,} {\bff #2}, #3}
\newcommand{\prltwoDE}[3]{ 20#1, {\itt Phys. Rev. Letts,} {\bff #2}, #3}
\newcommand{\revgeospphyDE}[3]{ 19#1, {\itt Rev. Geophys and Sp. Phys.,}
   {\bff #2}, #3}
\newcommand{\rppDE}[3]{ 19#1, {\itt Rep. Prog. Phys.,} {\bff #2}, #3}
\newcommand{\rpptwoDE}[3]{ 20#1, {\itt Rep. Prog. Phys.,} {\bff #2}, #3}
\newcommand{\ssrDE}[3]{ 19#1, {\itt Space Sci. Rev.,} {\bff #2}, #3}
\newcommand{\ssrtwoDE}[3]{ 20#1, {\itt Space Sci. Rev.,} {\bff #2}, #3}
\newcommand{\scienceDE}[3]{ 19#1, {\itt Science,} {\bff #2}, #3} 
\newcommand{\spDE}[3]{ 19#1, {\itt Solar Phys.,} {\bff #2}, #3} 
\newcommand{\spuDE}[3]{ 19#1, {\itt Sov. Phys. Usp.,} {\bff #2}, #3} 

\vskip-24pt\hbox{}

\end{document}